# Isotope characterization of shallow aquifers in the Horombe region, South of Madagascar

L.P. Fareze, J. Rajaobelison, V. Ramaroson, Raoelina Andriambololona,G. Andriamiarintsoa
*Madagascar-I.N.S.T.N., P.O. Box 4279 Antananarivo Madagascar*
P. R. Razafitsalama
*Projet d'Alimentation en Eau Potable et Assainissement, Ministère de l'Eau*
J.J. Rahobisoa, H. Randrianarison, A. Ranaivoarisoa
*Département de Science de la Terre, Faculté des Sciences, Université d'Antananarivo*
H. Marah
*Centre National de l'Energie, des Sciences et Techniques Nucléaires; Rabat (Morocco)*

The present study deals with the problem of evaluation of the recharge mechanism and the characterization of the groundwater flow system in the basement shallow aquifer, which is one of the groundwater resource in the semi-arid South region of Madagascar. Stable isotopes (deuterium and oxygen-18) and tritium are used to achieve with accuracy the hydrogeological and geochemical dynamics study. Chemical analysis is used to provide complementary information to the investigation. A space distribution of tritium concentration and isotopic composition in groundwater shows evidence of two opposite categories of aquifers, which is confirmed by the chemical analysis results and by the geological features of the study site. Some groundwater flow path directions have been identified in the study area thanks to the tritium concentration space distribution and the geological formation. Besides, the groundwater recharge of the shallow aquifers in the South of Madagascar has been characterized by the exponential mixing model.

## 1. INTRODUCTION

Isotope hydrology is a technique which calls for the use of isotopes to the water resources management. Since 1990, it became an important tool for investigating groundwater system and the aquifer vulnerability to contaminant. Environmental isotopes include stable isotopes and radioactive isotopes. Stable isotopes, which are invariant over time, are used to identify the origin of groundwater and mixing processes. Radioactive isotopes, whose decay over time can be detected, are used to assess the groundwater dynamics. Chemical analysis generally provides complementary information and is carried out in parallel with the isotopic investigation [2].

In 2006, a project, untitled "Support of isotope techniques to hydrogeological investigation for the National 700 boreholes drilling programme in the Fianarantsoa and Toliara Provinces", was launched in collaboration with IAEA. The main objective of this project was to apply isotopic and chemical tracing techniques for evaluating recharge rate, groundwater flow system, chemical characteristics of groundwater and aquifer vulnerability to pollution for the adequate integrated water resources management.

Preliminary results of this investigation are exploited to achieve the objective of the present study. Deuterium and oxygen-18 in groundwater and rainfall were used to determine the flow patterns in shallow aquifer in this region and mixing processes in groundwater [3]. Tritium was used to measure the age of groundwater in the shallow aquifer of the study area [4]. Thus, the specific objectives of this study are to define the available aquifers on site and contributes the groundwater flow paths directions by using isotope tracers. The identification of groundwater flow path allows predicting the risk of contamination into the aquifer [5].

## 2. DESCRIPTION OF THE STUDY AREA

### 2.1. Geological setting

The geological formation in Madagascar is constituted mainly of a crystalline basement of Precambrian age partially covered by a sedimentary layer (about 550M years to the quaternary). This basement covers the two-third of the island and the remaining one-third is mainly located in the west and covered by sedimentary layers.

The zone of survey is marked by a major shearing, the shearing of Bongolava-Ranotsara (N140° direction in N150°, sinistral direction), which disappears Westwards under the sedimentary layer. This shearing merges with a fault in the same direction which extends it to the East coast. It reaches a length of 350 km. Several zones of minor shearing and faults are also observed.

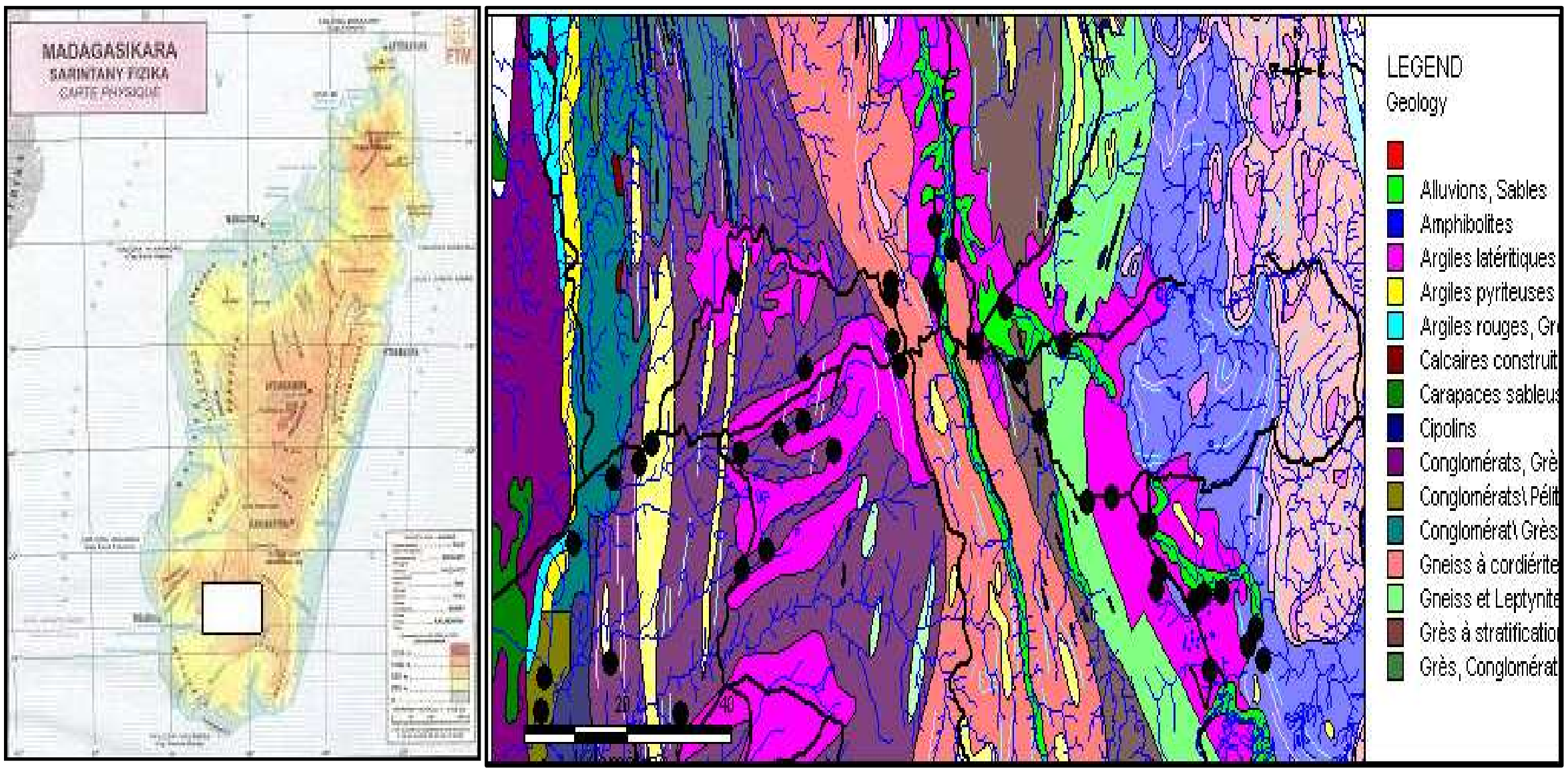

Figure 1: Geological map and sampling site of the study area

## 3. METHODOLOGY

### 3.1. Sampling

Sampling has been carried out in the Southern part of the region of Horombe by the Madagascar-INSTN team in collaboration with AEPA project (Ministry of Water). Samples have been collected during the winter season 2008 in 126 sites.

### 3.2. Analysis

Stable isotopes analyses were performed at CNESTEN (Rabat-Morocco). Chemical analysis and tritium measurements have been carried out at MADAGASCAR-INSTN. Tritium analyses were performed by electrolytic enrichment and liquid scintillation counting, and chemical analysis were performed by Ion Chromatography.

## 4. RESULTS AND INTERPRETATION

### 4.1. Stable Isotopes

The equation of Local Meteoric Water Line (Craig, 1961) is $\delta D = 8.058*\delta^{18}O + 14$. The samples isotopic compositions plot along two distinct lines characterised by equations $\delta D = 6.365*\delta^{18}O - 1.762$ and $\delta D = 7.579*\delta 18O + 5.84$ (figure 2), showing evidence of two distinct aquifers geologically separated by the Ranotsara fault . The results suggest as well that the aquifers are to some extent directly recharged by rainfalls, but somehow evaporated along the groundwater flow paths.

.

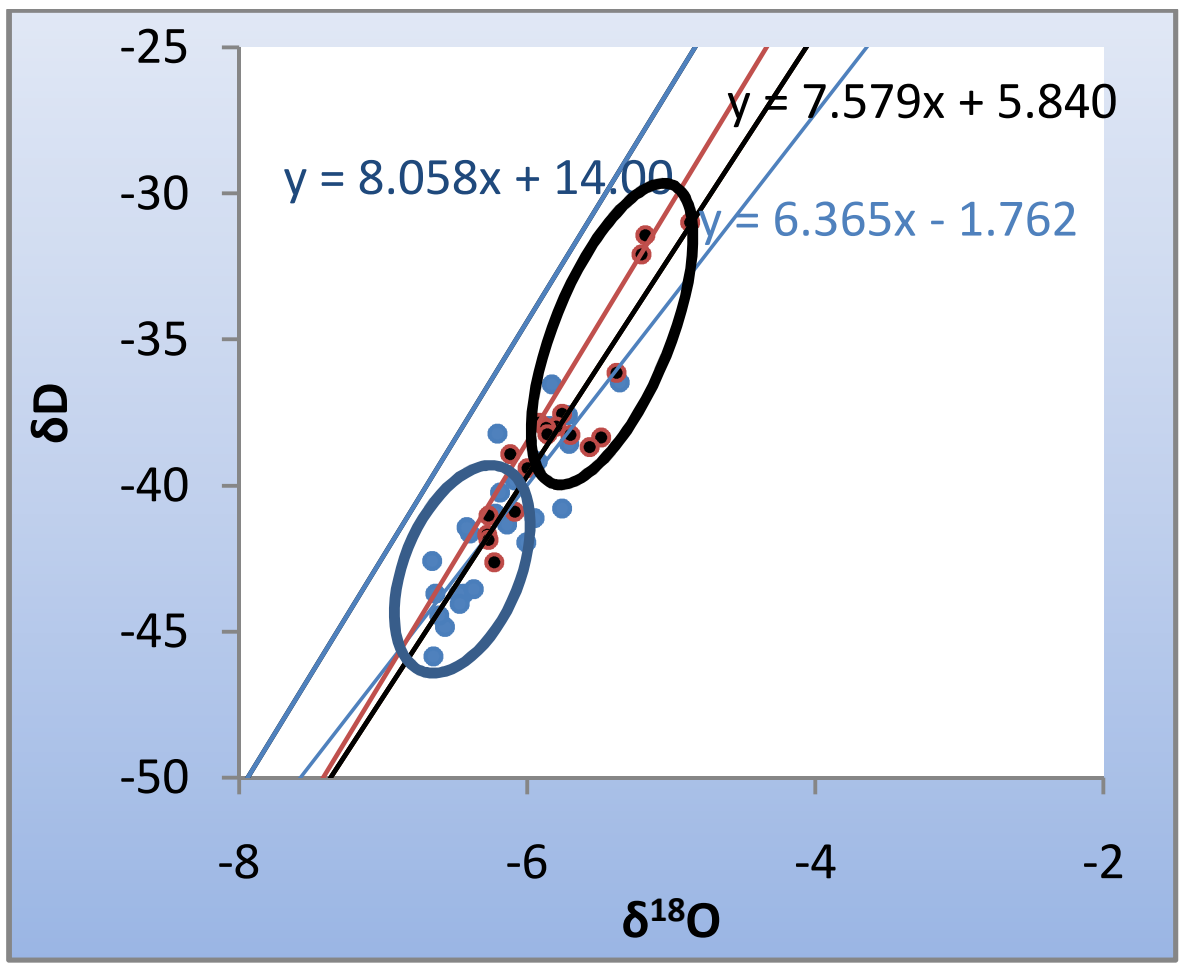

Figure 2: Plot of isotopic composition $\delta^{18}O$ and $\delta^{2}H$ of the groundwater samples in the study area

## 4.2. Tritium

The tritium concentration values in groundwater samples in the basement range from 0 TU to 2.97 TU (figure 6). The tritium results highlight the two opposite groups of aquifers which are evidenced by the stable isotopes results. In the western part of Ihosy River, the tritium values are relatively low (<0.1 TU), indicating the presence of relatively old groundwater in this part (Group B). However, replenishment occurs in the Eastern part of the River (Group A). The mean groundwater residence time, calculated from the input response curve in Pretoria (South Africa) between 1955 and 1985 and following an exponential model, is estimated to 80 years, which corresponds to a recharge rate of $11 mm.a^{-1}$ with an accuracy of about 30 % [1].

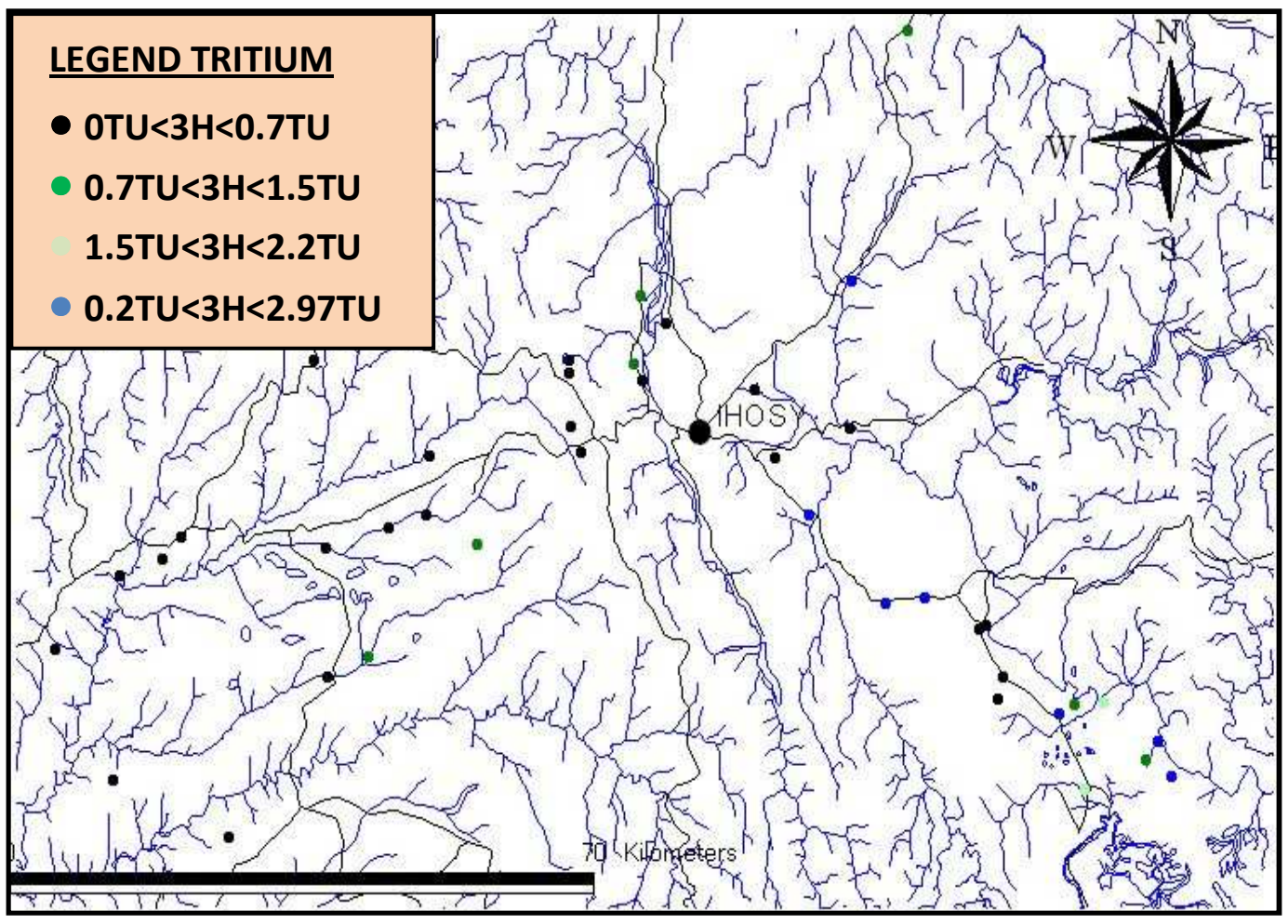

Figure 3: Spatial variation of groundwater tritium concentrations in the study area

## 4.3. Chemical analysis

The Piper diagram (Piper, 1944) shows two different chemical water types (figure 4a) : a sodium-calcium-bicarbonate water type characteristic of more mineralised ground water on one hand, and sodium-bicarbonate water type showing evidence of less mineralised fresh water. This result confirms the existence of the two opposite categories of aquifers which is clearly confirmed by the University of Avignon software plot (figure 4b). The groundwater on the left hand side of the fault of Ranotsara (Group B) is more mineralized probably because of their long residence time in the aquifer.

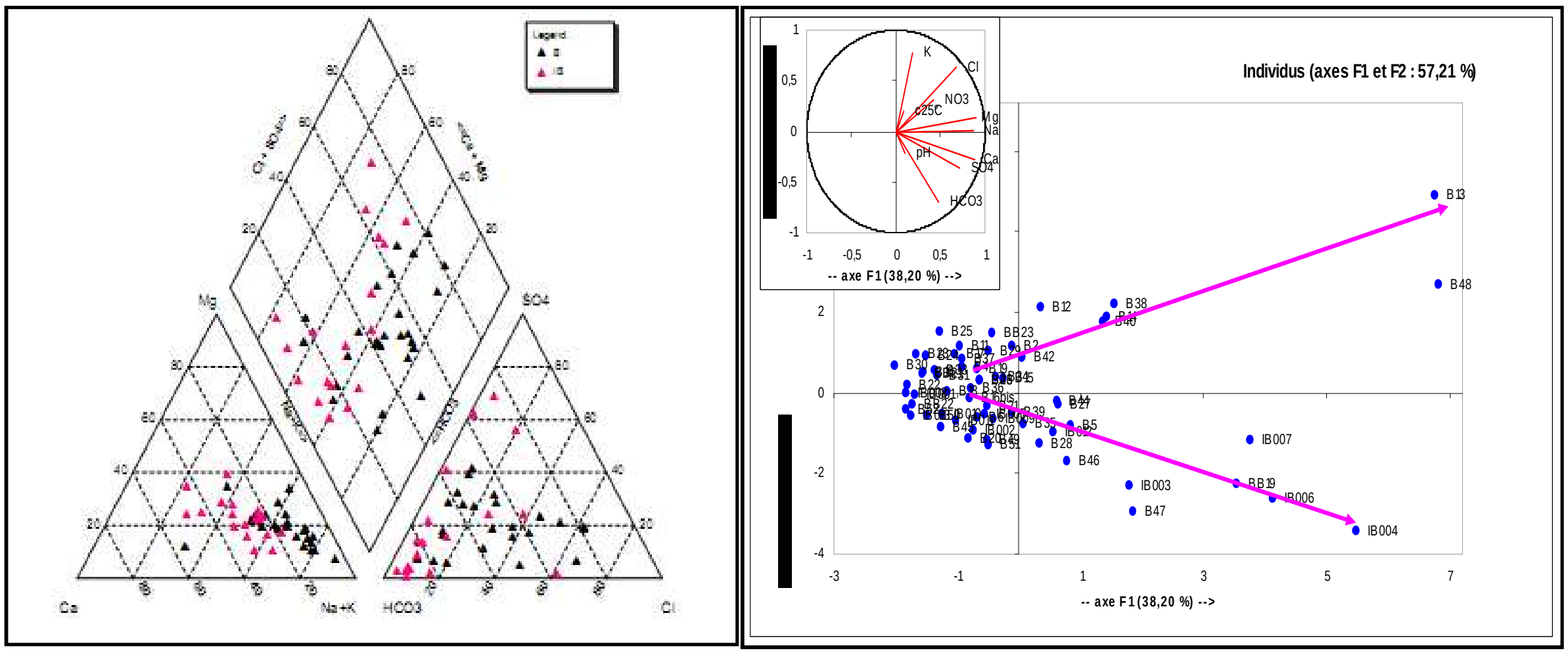

Figure 4: Piper diagram of ground water samples (left), Variable and individuals diagram related to ground waters (University of Avignon software) (right)

## 4.4. Aquifers vulnerability

The aquifer vulnerability may be expressed by the following equation:

$$Vulnerability = K * \frac{Tritium[TU]}{Transmissivity[m.s^{-1}] * Water\ level[m]}$$

Where, K is an arbitrary constant.

The vulnerability map is in accordance with the tritium map, confirming that the aquifers in Group A are more vulnerable to contamination because of its higher recharge than in GroupB.

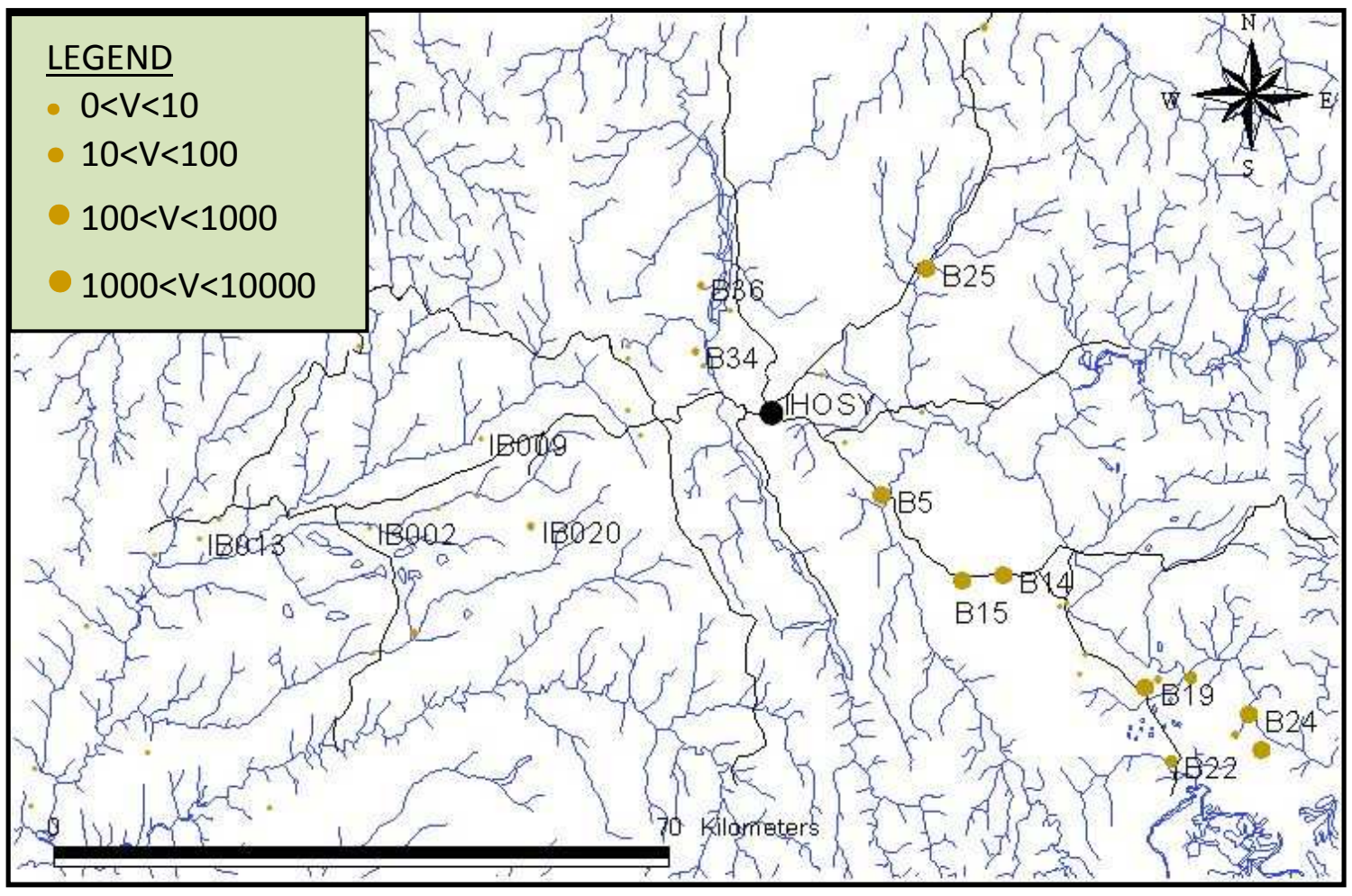

Figure 5: Vulnerability map of the study area

However, the vulnerability map does not cover the totality of the site water points because of the transmissivity data lack.

## 4.5. Ground water flow paths

Since tritium concentration decreases along the ground water flow paths, three directions of flow paths can be observed in the study area (figure 5). The first one starts in the Zazafotsy plain and moves South-Westwards up to Ihosy River. The second one follows the fault of Ranotsara in two opposite ways, South-Eastwards and North-Westwards (figure 5).

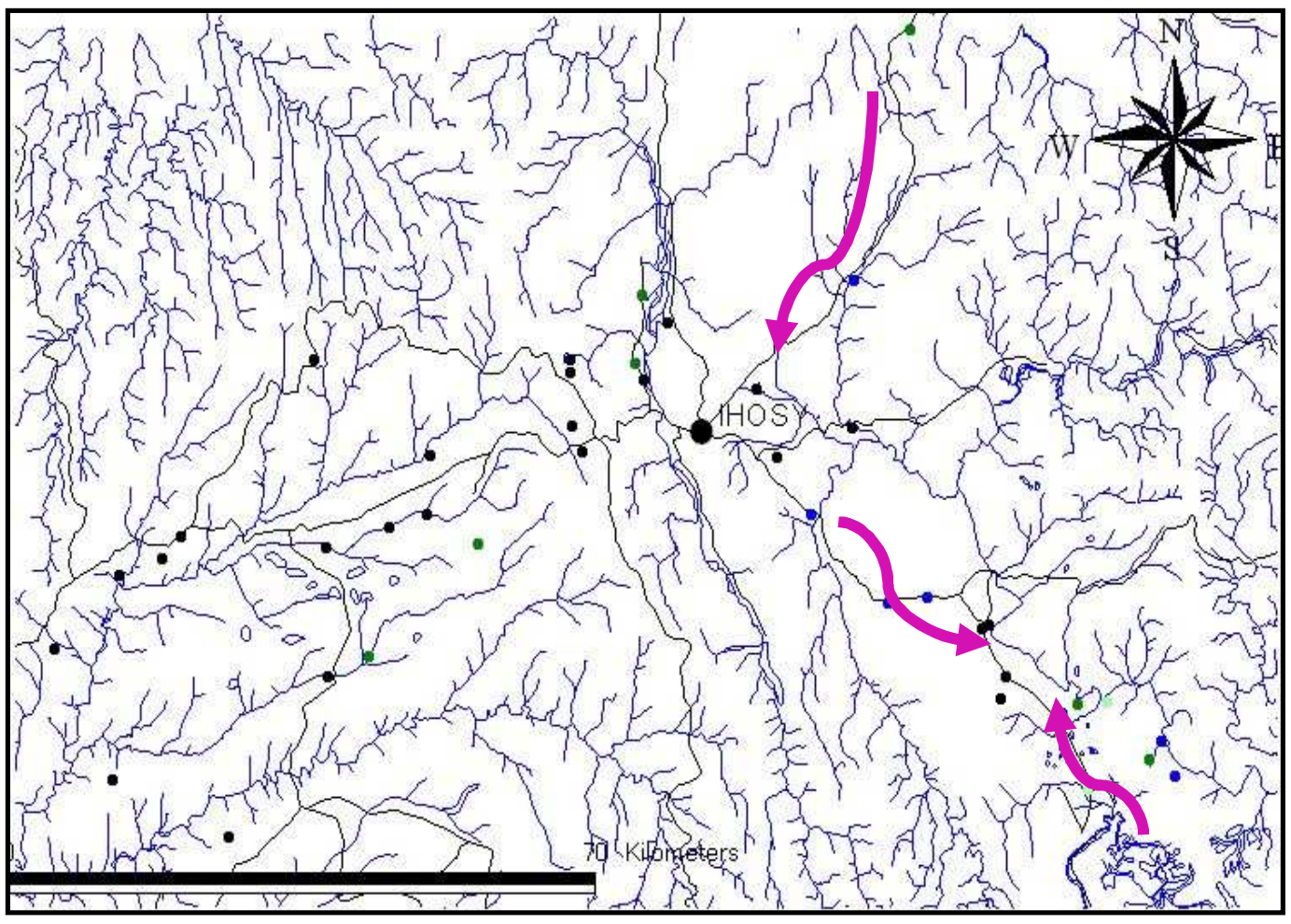

Figure 6: Groundwater flow paths in the shallow aquifer located in the Eastern part of the Ranotsara fault.

## 5. CONCLUSION

In the present case study, tritium and stable isotopes have proven to be powerful tools to determine the different types of aquifers, and their main hydrodynamic features (recharge rate, ground water flow path directions).

The results of the investigation allow us to conclude that the basement of this study area is divided in two opposite distinct aquifers. The first group (A), which is recharged, is located in the Eastern part of the Ranotsara fault, and the second one (B) which is somehow stagnant is located in the Western part of the latter fault. We can infer from these results that the Ranotsara fault is to some extent responsible for the hydrogeological features differences between group A and group B aquifers.

The plot of the stable isotopes composition shows that the aquifers, in the two groups of aquifers, are to some extent directly recharged by rainfalls, but slightly evaporated.

The groundwater in the south-western part of the Ihosy River (Group B) is more mineralized, probably because of a less important recharge rate in this part, proved by the very low concentration of tritium.

Moreover, the tritium concentrations spatial variation shows three trajectories of flow paths. The first one starts in the Zazafotsy plain towards the South-West direction, whereas the two other flow paths are located along the Ranotsara fault flowing in two opposite directions, the first one in the South East direction and the second one in the North West direction, both of them joining the Menarahaka River.

Consequently, the recharged shallow aquifer located in the East of the Ranotsara fault is more vulnerable to contamination.

## Acknowledgements

Our sincere gratitude to the International Atomic Energy Agency for its support within the frame of training, expertise and equipment, and to the Governmental authorities for their administrative and financial contributions.